\documentclass[aps,prd,preprintnumbers,nofootinbib,floatfix,11pt]{revtex4-2}
\pdfoutput=1
\usepackage{amsmath}
\usepackage{amssymb}
\usepackage{graphicx}
\usepackage{siunitx}
\usepackage{physics}
\usepackage{geometry}
\usepackage{hyperref}
\usepackage{caption}
\usepackage{subcaption}
\usepackage{tikz}
\usepackage{pgfplots}
\pgfplotsset{compat=1.16}
\usepgfplotslibrary{fillbetween}
\usepackage{braket}
\usepackage{xcolor}
\usepackage{comment}
\usepackage{nameref}
\usepackage{amsthm}

\usepackage[normalem]{ulem}

\begin{document}
\title{Double Resonance Strategy for Interferometric Detection of Axions}
\author{Spencer Green$^{1}$, Frank Wilczek$^{2,3,4,1}$} 
\affiliation{$^{1}$Center for Theoretical Physics, Massachusetts Institute of Technology, Cambridge, Massachusetts 02139, USA}
\affiliation{$^{2}$Department of Physics, Arizona State University, Tempe, Arizona 85287, USA}
\affiliation{$^{3}$Department of Physics, Stockholm University, Stockholm SE-106 91, Sweden}
\affiliation{$^{4}$Wilczek Quantum Center, SJTU and USTC, Shanghai 200240, China}

\date{\today}


\begin{abstract}
We propose a double-resonant interferometric strategy for axion dark matter detection that combines microwave circuit resonance with Fabry--P\'erot optical enhancement. In a strong magnetic field, axion--photon mixing induces a weak oscillating electric field, which is first amplified by a resonant circuit and then transduced into an optical phase shift via the electro-optic effect. Multiple coherent optical passes through the electro-optic medium accumulate this phase shift, enabling interferometric readout using mature optical techniques. We present the basic operating principle, discuss material requirements, and estimate the achievable sensitivity. For representative parameters, the projected reach extends into the parameter space of well-motivated QCD axion models.
\end{abstract}

\maketitle


Validation or disproof of the axion dark matter hypothesis remains one of the central experimental challenges in fundamental physics \cite{PDG2023, Ringwald24, Semertzidis22}. A wide variety of detection strategies are under active development, most prominently resonant microwave cavity experiments that attempt to extract axion-induced electromagnetic power \cite{Sikivie21, Rybka24}. Here we develop an alternative hybrid approach bringing in interferometric phase readout \cite{ebadi2024}. The essential idea is to employ \emph{double resonance}: a microwave resonator is used to enhance the axion-induced electric field generated via axion--photon mixing in a strong magnetic field, while an optical Fabry--P\'erot cavity coherently amplifies the resulting electro-optic phase modulation. This two-stage transduction---from axion field to resonantly enhanced electric field, and from electric field to optical phase---yields a sensitive and tunable probe of axions in the 10--40~GHz mass range and beyond.

In the presence of a strong magnetic field $B_0$, axions mix with photons according to the Lagrangian:
\begin{equation}
\mathcal{L}_{a\gamma\gamma} = -\frac{1}{4}g_{a\gamma\gamma}a F_{\mu\nu}\tilde{F}^{\mu\nu}
\end{equation}
where $g_{a\gamma\gamma}$ is the axion-photon coupling constant and $a$ is the axion field.

The mixing produces an effective oscillating electric field with amplitude:
\begin{equation}
E_0 = g_{a\gamma\gamma} B_0 \rho_a
\end{equation}
where $\rho_a = \sqrt{2\rho_{\text{DM}}\hbar c^3/(m_a c^2)}$ is the local axion field amplitude, with $\rho_{\text{DM}} \approx \SI{0.45}{\giga\electronvolt\per\centi\meter\cubed}$ being the dark matter density.

\section{Concept} 

For the sake of concreteness we will discuss a simple $LC$ circuit design, although alternatives, notably including split ring resonators, are also potentially interesting.  
The axion-induced electric field creates an oscillating voltage across a capacitor with capacitance:
\begin{equation}
C = \epsilon_r \epsilon_0 \frac{A}{d}
\end{equation}
where $\epsilon_r$ is the relative permittivity of the EO material, $A$ is the plate area, and $d$ is the gap.  At resonance $\omega = 1/\sqrt{LC}$, the electric field is enhanced by the quality factor:
\begin{equation}
E_{\text{enhanced}} = Q_{\text{LC}} \cdot E_0
\end{equation}
The LC quality factor is limited by dielectric losses of the EO material:
\begin{equation}
Q_{\text{LC}} = \frac{1}{\tan\delta} \approx 10^3
\end{equation}
for high-quality dielectrics at cryogenic temperatures.

The enhanced electric field modulates the refractive index of an electro-optic material \cite{Pockels} according to
\begin{equation}
\Delta n = \frac{1}{2}n^3 r_{33} E_{\text{enhanced}}
\end{equation}
where $n$ is the refractive index and $r_{33}$ is the relevant electro-optic coefficient (Pockels coefficient).
A single pass through the material produces a phase shift:
\begin{equation}
\Delta\phi_{\text{single}} = \frac{2\pi}{\lambda} \Delta n \cdot l = \frac{\pi n^3 r_{33} E_{\text{enhanced}} l}{\lambda}
\end{equation}

The Fabry-Pérot cavity provides multiple passes through the electro-optic material. The effective number of passes is:
\begin{equation}
N_{\text{eff}} = \frac{\mathcal{F}}{\pi} = \frac{Q_{\text{FP}}}{\pi} \cdot \frac{\lambda}{2l}
\end{equation}
where $\mathcal{F}$ is the finesse and $Q_{\text{FP}}$ is the optical cavity quality factor.   The total phase shift is therefore
\begin{equation}
\Delta\phi_{\text{total}} = N_{\text{eff}} \cdot \Delta\phi_{\text{single}} = \frac{Q_{\text{FP}} n^3 r_{33} E_{\text{enhanced}} l}{2}
\end{equation}

Conceptually, the detector operates as a two-stage transducer. First, in the presence of a strong static magnetic field, the axion field drives an oscillating electromagnetic response via axion--photon mixing. A microwave resonator converts this response into a spatially concentrated oscillating electric field whose amplitude is enhanced by the circuit quality factor. Second, this electric field modulates the refractive index of an electro-optic medium placed inside a Fabry--P\'erot cavity. A probe laser traversing the cavity acquires a phase modulation that is coherently amplified by multiple optical passes and read out interferometrically. The axion signal thus appears as a narrowband optical phase oscillation, with sensitivity set by the combined microwave and optical resonant enhancements.

It is useful to contrast the present strategy with other axion detection approaches. Conventional haloscope experiments \cite{ADMX, HAYSTAC} with metamaterial enhancements \cite{ALPHA23}, are designed to extract axion-induced electromagnetic \emph{power} from a resonant cavity and detect it with microwave receivers, with sensitivity scaling set by the achievable cavity volume, quality factor, and noise temperature. Dielectric haloscopes \cite{MADMAX, Millar2017} and related concepts similarly aim to enhance axion--photon conversion into propagating electromagnetic modes through impedance matching or coherent emission. By contrast, the scheme proposed here does not attempt to detect axion-induced microwave power directly. Instead, the axion field is first transduced into a resonantly enhanced, spatially concentrated oscillating electric field, which is then read out as an optical \emph{phase shift} using interferometric techniques proposed in \cite{ebadi2024}. This separation between field generation and signal readout enables the use of high-finesse optical cavities and mature phase-sensitive measurement methods, and naturally complements existing power-based searches, particularly in the tens-of-gigahertz mass range.  

\bigskip

\section{Tuning}

Efficient operation of the proposed detector requires simultaneous tuning of both the microwave resonator and the optical Fabry--P\'erot cavity. The axion field acts as a narrowband, quasi-monochromatic source with a frequency set by the axion mass and a finite coherence time determined by the dark matter velocity distribution. To achieve coherent signal enhancement, the microwave resonator must be tuned to the axion frequency to maximize the induced electric field, while the optical cavity must be configured such that the probe laser samples the electro-optic medium at a fixed phase of the axion-driven refractive index modulation.

The Fabry--P\'erot cavity length can be tuned via piezo-actuated mirror positioning, allowing continuous adjustment of the optical resonance condition. Because the axion-induced refractive index modulation oscillates at the axion Compton frequency, coherent optical phase accumulation requires that the cavity bandwidth be sufficient to track this modulation without averaging it away. For a cavity with finesse $\mathcal{F}$, this condition sets the characteristic frequency resolution over which the axion-induced signal remains coherent within the cavity,
\begin{equation}
\Delta f_{\mathrm{FP}} \sim \frac{m_a}{2\pi \mathcal{F}}.
\end{equation}

The microwave resonator must be tuned across the same frequency range to ensure overlap with the axion signal. For an LC circuit with quality factor $Q_{\mathrm{LC}}$, the resonance bandwidth is
\begin{equation}
\Delta f_{\mathrm{LC}} = \frac{f}{Q_{\mathrm{LC}}},
\end{equation}
which determines the frequency step size required to adequately scan axion mass. In practice, tuning may be achieved by varying the inductance or capacitance continuously, or by mechanically adjusting discrete resonant structures such as split-ring resonators.

Crucially, both resonances are continuously tunable, allowing systematic scans over axion mass without redesigning the apparatus. Once the two resonances are aligned, the axion-induced signal can be integrated coherently over the axion coherence time $\tau_c$, after which stochastic phase diffusion limits further coherent accumulation. As a result, for total integration time $T \gg \tau_c$, the signal-to-noise ratio grows as $(\tau_c T)^{1/4}$, as discussed below.

\bigskip

\section{Performance} 

We now estimate the performance achieved following this design, for the representative parameters summarized in Table 1.  

The axion-induced electric field in the presence of a magnetic field $B_0$ is:
\begin{equation}
E_a = g_{a\gamma\gamma} a B_0
\end{equation}
where $a$ is the axion field amplitude given by:
\begin{equation}
a = \sqrt{\frac{2\rho_{\text{DM}}\hbar c^3}{m_a c^2}}
\end{equation}

For our system parameters:
\begin{align}
\rho_{\text{DM}} &= \SI{0.45}{\giga\electronvolt\per\centi\meter\cubed} = \SI{7.2e-5}{\joule\per\meter\cubed} \\
B_0 &= \SI{10}{\tesla} \\
m_a &= \SI{100}{\micro\electronvolt} \quad \text{(at $\approx$25 GHz)}
\end{align}
The axion field amplitude is:
\begin{equation}
a \approx \SI{5.6e-13}{\volt} \times \left(\frac{\SI{100}{\micro\electronvolt}}{m_a}\right)^{1/2}
\end{equation}
Thus, the axion-induced electric field is:
\begin{equation}
E_a \approx \SI{5.6e-9}{\volt\per\meter} \times \left(\frac{g_{a\gamma\gamma}}{10^{-10}\text{ GeV}^{-1}}\right) \times \left(\frac{B_0}{\SI{10}{\tesla}}\right) \times \left(\frac{m_a}{\SI{100}{\micro\electronvolt}}\right)^{-1/2}
\end{equation}

A key innovation of our approach is the LC circuit enhancement. The electric field in the capacitor gap is enhanced by the LC circuit quality factor:
\begin{equation}
E_{\text{enhanced}} = \left(\frac{Q_{\text{LC}}}{10^3}\right) \cdot E_a
\end{equation}

The refractive index change in the electro-optic material is:
\begin{equation}
\Delta n = \frac{1}{2}n^3 r_{33} E_{\text{enhanced}}
\end{equation}
For LiNbO$_3$ with $n = 2.3$ and $r_{33} = 30 \times 10^{-12}$ m/V:
\begin{equation}
\Delta n = \frac{1}{2} \times 2.3^3 \times 30 \times 10^{-12} \times E_{\text{enhanced}} \approx 1.8 \times 10^{-10} \times E_{\text{enhanced}}
\end{equation}
The phase shift per pass through the material is:
\begin{equation}
\Delta\phi_{\text{single}} = \frac{2\pi}{\lambda} \Delta n \cdot l = \frac{\pi n^3 r_{33} E_{\text{enhanced}} l}{\lambda}
\end{equation}
With Fabry-Pérot enhancement, the total phase shift becomes:
\begin{equation}
\Delta\phi_{\text{total}} = N_{\text{eff}} \cdot \Delta\phi_{\text{single}} = \frac{Q_{\text{FP}} n^3 r_{33} E_{\text{a}} l}{2}
\end{equation}
For our parameters at \SI{25}{\giga\hertz}:
\begin{equation}
\Delta\phi_{\text{total}} \approx \SI{5e-15}{\radian} \times \left(\frac{g_{a\gamma\gamma}}{10^{-10}\text{ GeV}^{-1}}\right)
\end{equation}

\bigskip

\section{Signal and Noise}

The interferometer signal power is related to the phase shift by:
\begin{equation}
\delta P_{\text{out}} = \frac{2\pi}{\lambda} P_{\text{in}} N_{\text{eff}} \Delta\phi_{\text{single}}
\end{equation}
The shot-noise-limited amplitude spectral density is:
\begin{equation}
\text{ASD}_{\text{s.n.}} = \sqrt{2\hbar\omega_L P_{\text{out}}} = \sqrt{\hbar\omega_L P_{\text{in}}}
\end{equation}
For coherent oscillations over time $\tau_c$ and total measurement time $T$, the signal-to-noise ration (SNR) scales as:
\begin{equation}
\text{SNR} = \frac{\delta P_{\text{out}}}{\text{ASD}_{\text{s.n.}}} (\tau_c T)^{1/4}
\end{equation}
Where the coherence time is:
\begin{equation}
\tau_c \sim \frac{1}{m_a v^2} \approx \frac{10^6}{m_a[\mu\text{eV}]} \text{ s}
\end{equation}

For our system with LiNbO$_3$:
\begin{eqnarray}
\text{SNR} \approx& \nonumber \\
&\left(\frac{L_0 N_{\text{eff}}Q_{LC}}{\SI{10}{\milli\meter} \times 1.5 \times 10^5}\right) \left(\frac{\lambda}{\SI{1064}{\nano\meter}}\right)^{-1/2} \left(\frac{P_{\text{in}}}{\SI{5}{\watt}}\right)^{1/2} \left(\frac{T}{\text{s}}\right)^{1/4} \left(\frac{g_{a\gamma\gamma}}{10^{-10}\text{ GeV}^{-1}}\right) \left(\frac{m_a}{\SI{100}{\micro\electronvolt}}\right)^{-5/4}
\end{eqnarray}

\section{Sensitivity Relative to Models}

The two reference QCD axion models predict different relationships between the axion mass and coupling. For the KSVZ model we have
\begin{equation}
g_{a\gamma\gamma}^{\text{KSVZ}} = \frac{\alpha}{2\pi f_a} \times 0.97 = \frac{m_a}{5.7 \times 10^6 \text{ eV}} \times 1.1 \times 10^{-15} \text{ GeV}^{-1}
\end{equation}
while for the DFSZ model we have
\begin{equation}
g_{a\gamma\gamma}^{\text{DFSZ}} = \frac{\alpha}{2\pi f_a} \times 0.36 = \frac{m_a}{5.7 \times 10^6 \text{ eV}} \times 4.0 \times 10^{-16} \text{ GeV}^{-1}
\end{equation}

The axion mass is related to the decay constant as:
\begin{equation}
m_a = \frac{5.7 \times 10^{6} \text{ eV}}{f_a/\text{GeV}}
\end{equation}

Using a $10 \text{ T}$ magnet and $10 \text{ W}$ laser, our calculations indicate the proposed experiment could reach the KSVZ band with only $1\, \sec$ integrations per packet, and the DFSZ band with $100\, \sec$ for $\text{SNR}=1$.

\begin{table}
\centering
\begin{tabular}{|l|c|c|}
\hline
\textbf{Parameter} & \textbf{Value} & \textbf{Units} \\
\hline
Magnetic field ($B_0$) & 10 & T \\
Operating temperature & 4 & K \\
Laser wavelength & 1064 & nm \\
Laser Power ($P$) & 10 & W \\
EO material & LiNbO$_3$ & - \\
Refractive index ($n$) & 2.3 & - \\
EO coefficient ($r_{33}$) & $30 \times 10^{-12}$ & m/V \\
Loss tangent ($\tan\delta$) & $10^{-3}$ & - \\
\hline
\end{tabular}
\caption{Representative experimental parameters}
\end{table}

\begin{figure}
\centering{\includegraphics[width=6in]{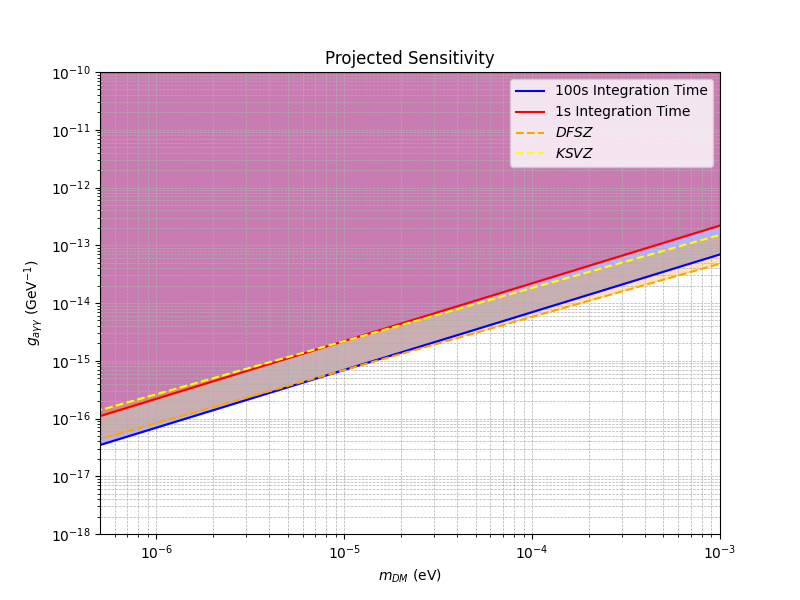}}
\caption{Projected sensitivity of experiment with $\text{SNR}=1$ and a $10 \text{ T}$ magnet, $10 \text{ W}$ laser, and integration times of (a) $100\sec$ (b) $1\sec$.}

\end{figure}

\section{Optimizations and Pitfalls}

In making our estimates we have chosen a relatively simple, easy-to-analyze geometry and parameters characteristic of readily available magnets, materials, and lasers.  All of these are plausibly open to further optimization.  Let us mention in particular the possibilities of employing split-ring resonators in place of simple LC circuits, and of constructing arrays in either configuration.  

We should also note some significant engineering challenges that will have to be investigated and addressed in mounting a practical experiment:
\begin{itemize}
\item Maintaining high $Q$ with participation of an electro-optic material
\item Avoiding optical heating of the slab at cryogenic temperatures
\item Avoiding noise injection from vibrations
\item Limiting microwave thermal and quantum noise in the resonator and coupling network
\end{itemize}



\section{Conclusion}

We have proposed a double-resonant interferometric strategy for axion dark matter detection in which the axion field is transduced in two stages: first into a resonantly enhanced oscillating electric field via axion--photon mixing in a microwave circuit, and then into an optical phase modulation via the electro-optic effect. The combination of microwave and optical resonance enables substantial coherent enhancement while retaining practical tunability across axion mass. The resulting interferometric readout leverages mature, phase-sensitive optical techniques rather than direct microwave power detection.

Our estimates, based on conservative parameters and a simple geometry, indicate that this approach can probe well-motivated QCD axion models in the tens-of-gigahertz mass range with modest integration times. While significant experimental challenges remain---including cryogenic operation with electro-optic materials, vibration isolation, and microwave noise control---the central scaling arguments appear favorable. We view the present work as establishing a promising new architecture for axion searches that is complementary to existing power-based experiments and opens several avenues for further optimization.

\section{Acknowledgments}
FW is supported by the Swedish Research Council under Contract No. 335-2014-7424.

\bibliographystyle{apsrev4-2}
\bibliography{refs}

@article{PDG2023,
  author       = {Workman, R. L. and others},
  collaboration= {Particle Data Group},
  title        = {Review of Particle Physics},
  journal      = {Prog. Theor. Exp. Phys.},
  volume       = {2022},
  pages        = {083C01},
  year         = {2022},
  doi          = {10.1093/ptep/ptac097},
  note         = {Cited in the paper as PDG 2023 update (see PDG online updates).}
}

@misc{Ringwald24,
  author       = {Ringwald, Andreas},
  title        = {Review on Axions},
  eprint       = {2404.09036},
  archivePrefix= {arXiv},
  primaryClass = {hep-ph},
  year         = {2024}
}

@article{Semertzidis22,
  author       = {Semertzidis, Yannis K. and Youn, SungWoo},
  title        = {Axion dark matter: How to see it?},
  journal      = {Sci. Adv.},
  volume       = {8},
  number       = {8},
  pages        = {eabm9928},
  year         = {2022},
  doi          = {10.1126/sciadv.abm9928}
}

@article{Sikivie21,
  author       = {Sikivie, Pierre},
  title        = {Invisible axion search methods},
  journal      = {Rev. Mod. Phys.},
  volume       = {93},
  pages        = {015004},
  year         = {2021},
  doi          = {10.1103/RevModPhys.93.015004},
  eprint       = {2003.02206},
  archivePrefix= {arXiv},
  primaryClass = {hep-ph}
}

@article{Rybka24,
  author       = {Rybka, Gray},
  title        = {Axion dark matter searches above 1 $\mu$eV},
  journal      = {Nucl. Phys. B},
  volume       = {1003},
  pages        = {116481},
  year         = {2024},
  doi          = {10.1016/j.nuclphysb.2024.116481}
}

@article{ebadi2024,
  author       = {Ebadi, Reza and Kaplan, David E. and Rajendran, Surjeet and Walsworth, Ronald L.},
  title        = {GALILEO: Galactic Axion Laser Interferometer Leveraging Electro-Optics},
  journal      = {Phys. Rev. Lett.},
  volume       = {132},
  pages        = {101001},
  year         = {2024},
  doi          = {10.1103/PhysRevLett.132.101001}
}

@book{Pockels,
  author       = {Boyd, Robert W.},
  title        = {Nonlinear Optics},
  publisher    = {Academic Press},
  year         = {2008}
}

@article{ADMX,
  author       = {Du, N. and others},
  collaboration= {ADMX Collaboration},
  title        = {Search for Invisible Axion Dark Matter with the Axion Dark Matter Experiment},
  journal      = {Phys. Rev. Lett.},
  volume       = {120},
  pages        = {151301},
  year         = {2018},
  doi          = {10.1103/PhysRevLett.120.151301},
  eprint       = {1804.05750},
  archivePrefix= {arXiv},
  primaryClass = {hep-ex}
}

@article{HAYSTAC,
  author       = {Zhong, L. and others},
  title        = {Results from Phase 1 of the HAYSTAC Microwave Cavity Axion Experiment},
  journal      = {Phys. Rev. D},
  volume       = {97},
  pages        = {092001},
  year         = {2018},
  doi          = {10.1103/PhysRevD.97.092001},
  eprint       = {1803.03690},
  archivePrefix= {arXiv},
  primaryClass = {hep-ex}
}

@article{ALPHA23,
  author       = {Millar, Alexander J. and others},
  collaboration= {ALPHA Collaboration},
  title        = {Searching for dark matter with plasma haloscopes},
  journal      = {Phys. Rev. D},
  volume       = {107},
  pages        = {055013},
  year         = {2023},
  doi          = {10.1103/PhysRevD.107.055013},
  eprint       = {2210.00017},
  archivePrefix= {arXiv},
  primaryClass = {hep-ph}
}

@article{MADMAX,
  author       = {Caldwell, Allen and others},
  title        = {Dielectric Haloscopes: A New Way to Detect Axion Dark Matter},
  journal      = {Phys. Rev. Lett.},
  volume       = {118},
  pages        = {091801},
  year         = {2017},
  doi          = {10.1103/PhysRevLett.118.091801},
  eprint       = {1611.05865},
  archivePrefix= {arXiv},
  primaryClass = {hep-ph}
}

@article{Millar2017,
  author       = {Millar, Alexander J. and Raffelt, Georg G. and Redondo, Javier and Steffen, Frank D.},
  title        = {Dielectric Haloscopes to Search for Axion Dark Matter},
  journal      = {Phys. Rev. D},
  volume       = {98},
  pages        = {123008},
  year         = {2018},
  doi          = {10.1103/PhysRevD.98.123008},
  eprint       = {1612.07057},
  archivePrefix= {arXiv},
  primaryClass = {hep-ph}
}

\end{document}